\documentstyle[aps]{revtex}
\begin{document}
\tightenlines
\draft
\title{Scalar field perturbations in Fresh inflation}
\author{Mauricio Bellini\footnote{E-mail address: mbellini@mdp.edu.ar}}
\address{Consejo Nacional de Investigaciones Cient\'{\i}ficas y
T\'ecnicas (CONICET)\\
and\\
Departamento de F\'{\i}sica,
Facultad de Ciencias Exactas y Naturales,
Universidad Nacional de Mar del Plata, \\
Funes 3350, (7600) Mar del Plata, Buenos Aires, Argentina.}
\maketitle
\begin{abstract}
The model of fresh inflation with increasing cosmological parameter
provides sufficient e-folds to solve the
flatness/horizon problem and the density fluctuations agree with
experimental values. In this model the
temperature increases during fresh inflation
and reach its maximum value when inflation ends where the $N$-number
of e-folds is $N(t_e) \gg 60$.
The most important characteristic of this model is that
provides a natural transition between the end of inflation and the
epoch when the universe is radiation dominated.
\end{abstract}
\vskip .2cm                             
\noindent
Pacs numbers: 98.80.Cq\\
\vskip 1cm
Inflation is one of the most reliable concepts in modern cosmology.
The first model of inflation was proposed by A. Starobinsky\cite{1.}.
A much simpler inflationary model with a clear motivation was developed
by A. Guth\cite{2.}, in order to solve some of the shortcomings of the
big bang theory, and in particular, to explain the extraordinary homogeneity
of the observable universe. However, the universe after inflation in this
scenario becomes very inhomogeneous. These problems were sorted out
by A. Linde in 1983 with the introduction of chaotic inflation\cite{3.}.
In this scenario inflation can occur in theories with potentials such as
$V(\phi) \sim \phi^n$. It may begin in the absence of thermal equilibrium
in the early universe, and may start even at the Planckian density, in which
case the problem of initial conditions for inflation can be easily
solved.
Although a justification from first principles for dissipative
effects has not been firmly achieved in the framework of inflation,
such effects should not be ruled out on the basis of
readiness alone. Much work can be done on phenomenological grounds as,
for instance, by applying nonequilibrium thermodynamic techniques
to the problem or even studying particular models with dissipation.
An interesting example of the latter case is the warm inflationary
picture\cite{1,1a}. As in new inflation\cite{s,s1}, a
phase transition driving
the universe to an inflationary period dominated by the scalar field
potential is assumed. A standard phenomenological frictionlike
term $\Gamma \dot\phi$ is inserted into
the scalar field equation of motion to represent a continuous
energy transfered from $\phi$ to the radiation field. This persistent thermal
contact during inflation is so finely adjusted that the scalar field
evolves all the time in a damped regime generating an isothermal expansion.
As a consequence, the subsequent reheating mechanism is not
needed and thermal fluctuations produce the primordial spectrum of
density perturbations\cite{2}. More recently was demonstrated
that isentropic and warm pictures are just extremme cases of an
infinite two-parametric family of possible inflationary scenarios\cite{ml}.
Very recently, a dissipative mechanism which emerges in generic interacting
quantum field systems and leads to robust warm inflation was developed\cite{b}.

Two years ago, a
new scenario called fresh inflation was introduced\cite{4,4a}.
It can be viewed as a unification of both chaotic and
warm inflation scenarios with constant\cite{4} and increasing\cite{4a}
$F$-cosmolgical parameter.
As in chaotic inflation the universe begins from an unstable
primordial matter field perturbation with energy density nearly
${\rm M}^4_p$ (${\rm M}_p =1.2 \times GeV$
is the Planckian mass) and chaotic
initial conditions. Furthermore, initially
the universe there is not thermalized
so that the radiation energy density when inflation starts is zero 
[$\rho_r(t=t_0)=0$].
It must to be point out that quantum field theory first principle
calculations of radiation generated from a zero temperature state
has been done in\cite{4aa}, but
this formalism is realized in an non-expanding Minkowski spacetime.

As initial time we understand the Planckian time
$G^{1/2}$, where $G$ is the gravitational constant.
Later, the universe will describe a second-order
phase transition. In other words, the inflaton rolls down towards the
minimum energetic configuration. 
Particle production and thermalization occur together during the rapid
expansion of the universe, so that the radiation energy density grows
during fresh inflation ($\dot\rho_r >0$).
The interaction between the inflation field and the particles produced
during inflation provides slow-rolling of the inflaton field.
So, in the fresh inflationary model (also in warm inflation), the
slow-roll conditions are physically well justified.
The decay width of the $\phi$-field grows with time, so when
the inflaton approaches to the minimum of the potential there
is no oscillation around the minimum energetic configuration. Hence,
the reheating period does not happen in fresh inflation.
This model attempts to build a bridge between the standard and
warm inflationary models, beginning from
chaotic initial conditions which provides naturality.
In fresh inflation with constant cosmological parameter the
universe can be seen as inflating in a $4D$ Friedmann-Robertson-Walker
(FRW) metric embedding in
a $5D$ metric in apparent vacuum when the fifth coordinate of the metric
is non-compact and remains constant\cite{GRG1}. Furthermore, can be demonstrated
the possibility that violation of baryon number conservation\cite{GRG2} and
ultralight particle creation\cite{NC1} can occur during the period
out-of-equilibrium, i.e., before the thermal equilibrium to be restored
at the end of fresh inflation.
Gauge-invariant metric fluctuations in a fresh inflationary model with
increasing cosmological parameter was considered in\cite{4a}. The main
aim of this paper is the study of the evolution of density perturbations
in such that model.

We describe fresh inflation with a
Lagrangian for a $\phi$-scalar field minimally coupled to gravity,
which also interacts with another $\psi$-scalar field by means of
${\cal L}_{int} = -{\rm g}^2 \phi^2 \psi^2$, is
\begin{equation}\label{1}
{\cal L} = \sqrt{-g}
\left[\frac{R}{16\pi G} - \frac{1}{2}\left(
\nabla \phi\right)^2 - V(\phi)+{\cal L}_{int}\right],
\end{equation}
where $R= 6(a \ddot a+\dot a^2)/a^2$
is the scalar curvature, $a$ is the scale factor of the universe and
$g$ is the determinant of the metric tensor $g^{\mu\nu}$ with
$\mu,\nu = 0,1,2,3$. In this paper I consider a
FRW metric for a spatially flat, isotropic and homogeneous universe
described by the line element $ds^2 = -dt^2 + a^2 dr^2$.
If $\delta = \dot\rho_r+ 4 H \rho_r$ describes the interaction between
the inflaton and the bath for a $\gamma=4/3$-fluid which
expands with a Hubble parameter $H=\dot a/a$ and
radiation energy density $\rho_r$, hence the
equations of motion for $\phi$ and radiation energy density are
\begin{equation}\label{rho}
\ddot\phi + 3 H \dot\phi + V'(\phi) + \frac{\delta}{\dot\phi}  = 0,
\quad
\dot\rho_r + 4 H \rho_r - \delta =  0.
\end{equation}
Here, $\delta = \Gamma(\theta) \dot\phi^2$ describes a Yukawa
interaction and the $\phi$-decay width is
$\Gamma(\theta) = [g^4_{\rm eff}/(192 \pi)] \theta$\cite{ber}.
Furthermore, $\theta \sim \rho^{1/4}_r$ is the temperature of the bath.
The cosmological parameter $F = (p_t + \rho_t)/\rho_t$ describes
the evolution of the universe during inflation
\begin{equation}\label{3}
F = -\frac{2 \dot H}{3 H^2} = \frac{\dot\phi^2 + \frac{4}{3} \rho_r}{\rho_r
+ \frac{\dot\phi^2}{2} + V} ,
\end{equation}
where the total pressure and
energy density are given respectively by
\begin{equation}
p_t  = \frac{\dot\phi^2}{2} + \frac{\rho_r}{3} - V(\phi), \quad
\rho_t = \rho_r + \frac{\dot\phi^2}{2} + V(\phi).
\end{equation}
In the previous works\cite{4} only was considered the case
where the cosmological parameter $F$ is a constant.
However, as we can see in eq. (\ref{3}), during inflation
the potential energy density decreases, so that the
radiation energy density becomes more important in $F$. This
means that $F$ must be increasing during fresh inflation, but
of course, always remaining below $4/3$, which corresponds to
a radiation dominated universe.
We can write $\rho_r$ and $V(\phi)$
as a function of $\phi$\cite{4}
\begin{equation}\label{p}
\rho_r = \left(\frac{3F}{4-3F}\right) V -
\frac{27}{8} \left(\frac{H^2}{H'}\right)^2 \frac{F^2(2-F)}{(4-3F)}, \quad
V(\phi) = \frac{3}{ 8 \pi G} \left[ \left(\frac{4-3F}{4}\right) H^2
+ \frac{3\pi G}{2} F^2 \left(\frac{H^2}{H'}\right)^2\right], 
\end{equation}
where $F$ is a function of $\phi$ and the time evolution of $\phi$ is
described by the equation
\begin{equation}\label{phi}
\dot\phi = - \frac{3 H^2}{2 H'} F(\phi).
\end{equation}
We consider in the second equation of (\ref{p}) the potential
$V(\phi) =[{\cal M}^2(0)/2]\phi^2+ [\lambda^2/4]\phi^4$, where
${\rm G}=M^{-2}_p$ is the gravitational constant
and $M_p=1.2 \  10^{19} \  {\rm GeV}$
is the Planckian mass.
The inflaton field is really an effective field described by
$\phi=(\phi_i \phi_i)^{1/2}$. Furthermore, ${\cal M}^2(0)$ is
given by ${\cal M}^2_0$ plus renormalization counterterms in the
initial potential ${1\over 2}{\cal M}^2_0 (\phi_i\phi_i)
+ {\lambda^2\over 4}(\phi_i\phi_i)^2$\cite{55}, the effective potential
is $V_{{\rm eff}}(\phi,\theta)
= [{\cal M}^2(\theta)/2]\phi^2+[\lambda^2/4]\phi^4$.
Here,
$\theta$ is the temperature
and ${\cal M}^2(\theta) = {\cal M}^2(0)+ {(n+2) \over 12} \lambda^2
\theta^2$, such that $V_{{\rm eff}}(\phi,\theta) =
V(\phi) + \rho_r(\theta,\phi)$.
The temperature increases
with the expansion of the universe because the inflaton transfers radiation
energy density to the bath with a rate larger than the expansion
of the universe.
So, the number of created particles $n$ [for
$\rho_r=(\pi^2/30) g_{\rm eff} \theta^4
= \left({\cal M}^2(\theta)-{\cal M}^2(0)\right)\phi^2$], is given by
\begin{equation}\label{n}
(n+2) = \frac{2\pi^2}{5\lambda^2} g_{{\rm eff}} \frac{\theta^2}{\phi^2},
\end{equation}
where $g_{{\rm eff}}$ denotes the effective degrees of freedom of the
particles and it is assumed that $\phi$ has no self-interaction.
Note that eq. (\ref{n}) only depends on the choice of $V_{{\rm eff}}$ and
${\cal M}^2(\theta)$, and not on $H$ and $F$.
(For a more detailed explanation of eq. (\ref{n}) the reader
can see the papers \cite{4}.)
The density fluctuacions are given by
\begin{equation}\label{18}
\frac{\left<\delta\rho_t\right>}{\rho_t} \simeq
\Delta(\phi) \left<\delta\phi^2\right>^{1/2},
\end{equation}
where $\delta\phi^2$ are the squared fluctuations of the inflaton field
and $\Delta(\phi)={V'_{eff} \over \dot\phi^2 + 4/3\rho_r}$.
The notation $<...>$ denotes the averaging with respect to a gaussian
distribution.
Since the temperature of the thermal bath is $\theta$, we can consider
the Langevin equation for the field fluctuations $\delta\phi$, with
a white and gaussian noise $\kappa(t)$, due to the stochastic interaction
of the field $\phi$ with the thermalyzed environment
\begin{equation}
\dot\delta\phi \simeq \frac{H^2}{3H +\Gamma} \delta\phi + \kappa(t).
\end{equation}
Here, we have made use of the fluctuation dissipation theorem\cite{RMP},
and the fact that the slow-roll condition holds.
This fact implies that $H^2 \gg V''$. 
The diffusion coefficient for the fluctuations with a wavelenght given by
the Hubble radius $H^{-1}$, is (see the first paper in \cite{2})
$ D = {3 \over 2\pi} {\theta H^3 \over 3H+ \Gamma}$,
such that the squared mean fluctuations of the field are\cite{BF}
\begin{equation}\label{22}
\left< \delta\phi^2 \right> \sim \sqrt{\Gamma H} \theta.
\end{equation}
This relation is only valid when the system is thermalyzed, i.e., in the
last stages of fresh inflation for which the relation $\Gamma > H$ holds.

As an example we
consider the particular case where the Hubble and cosmological
parameters are $ H(\phi) =  A \  \phi^2 $ and $ F(\phi) = B \  \phi^{-2}$,
being $A$ and $B$ ${\rm G}^{1/2}$ and
${\rm G}^{-1}$ dimensional constants which can be obtained by replacing
$H$ and $F$ in eq. (\ref{p})
\begin{equation}\label{a}
A^2  =  \frac{\lambda^2}{4} \frac{8 \pi G}{3}, \quad
B =  \frac{3 \lambda + \sqrt{9 \lambda^2
+ 48 \pi G {\cal M}^2(0)}}{3 \lambda
\pi G}. 
\end{equation}
From eq. (\ref{phi}) we obtain the time evolution for the inflaton field
$\phi(t) = \phi^{(s)} \  e^{- {3 AB \over 2} t}$,
where $\phi^{(s)}$ is its initial value. It must be sufficiently large to
assures at least,
the $50-60$ e-folds needed during inflation before transcurred
($10^{10}-10^{12}$) Planckian times. Since $H = \dot a/a$, the time
evolution of the scale factor will be
$a = a_0 \  e^{- {\phi^2(t)\over 3 B}}$,
which increases with time due to the decreasing of $\phi(t)$.
Furthermore, the temperature, written as a function of the field
is obtained from the second equation in (\ref{rho})
\begin{equation}
\theta(\phi) = \left\{4\pi \phi (8\phi - 3B)
\left[ 16 {\cal M}^2(0) + \phi^2 \left(8\lambda^2
- 18 A^2 B\right) +
9 A^2 B^2\right]\right\} \left[B A g^4_{\rm eff}\left(4\phi^2 - 3 B
\right)\right]^{-1},
\end{equation}
which is zero when fresh inflation starts. The $\phi$-value at this
time is $\phi^{(s)} = 3 B/8$.
The function $\Delta(\phi)$ in eq. (\ref{18}) is

\begin{equation}\label{19}
\Delta(\phi) = \frac{3 \left(\pi G B-2\right)}{4 \phi}.
\end{equation}
Hence, from eqs. (\ref{19}) and (\ref{22}) one obtains the
density fluctuations ${<\delta\rho_t >\over \rho_t}$ in eq. (\ref{18})
\begin{equation}
\frac{\left<\delta\rho_t\right>}{\rho_t} \sim
\frac{3(\pi B G -2)}{4\phi} 
\left[\frac{\pi \phi^3 (3 B - 8 \phi)}
{g^4_{\rm eff} B(4\phi^2-3 B)}\right]^{1/2} 
\left[2\phi^2 (9A^2 B-4\lambda^2)-16 {\cal M}^2(0)-9 A^2 B^2\right]^{1/2}.
\end{equation}

Figures (1), (2) and (3) 
show respectively the evolution of the temperature $\theta(t)$,
density fluctuations $<\delta\rho_t>/\rho_t$ (as a function
of the number of $e$-folds) and the
number of created particles in eq. (\ref{n}) (as a function of time).
These graphics were made using the values
${\cal M}(0)=1.5 \times 10^{-9} \  {\rm G}^{-1/2}$, $\lambda=4\times
10^{-15}$ and $g_{\rm eff}=25$. With these parameter values we obtain
respectively $A=0.578 \times 10^{-14} \  {\rm G}^{1/2}$ and
$B=488726.728 \  {\rm G}^{-1}$ in eqs. (\ref{a}).
Notice that fresh inflation ends at $t_e\simeq 2\times 10^8 \  {\rm G}^{1/2}$,
when $\theta(t_e) \simeq 6\times 10^{-8} \  {\rm G}^{-1/2}$
[see Fig. (1)] and $\left.<\delta\rho_t>/\rho_t\right|_{N(t_e)} \simeq
10^{-5}$ [see
Fig. (2)] reach their maximum values.
The $\phi$-value when fresh inflation ends is given by $\phi^{(e)}\equiv
\phi(t_e) \simeq 8 \times 10^5 \  {\rm G}^{-1/2}$.
Note that $N(t_e) \gg 60$ at this moment so that the model here studied
solve the flatness/horizon problems.
Furthermore,
$\phi$ takes transplanckian values
during fresh inflation with increasing cosmological
parameter. It could be a problem for this model because there is no
clear the validity of conventional quantum field theory in the
transplanckian regime.

Finally, Fig. (3) shows the evolution of the number
of created particles, which grows to reach its asymptotic value
$n \simeq 9 \times 10^6$, at the end of the inflationary period.

To summarize,
the conditions on $\rho_r$ lead to two different types of thermodynamic
regimes of inflation. At the begining $\rho_r \simeq 0$
and the expansion is isentropic during
inflation, but the temperature increases
because
$\Gamma$ is sufficietly strong to make $\dot\rho_r >0$.
This is the case of fresh inflation, in which this first regime is
followed by a non-isentropic period in which
$\rho_v >\rho_r >0$, so that the
temperature of the universe may still be sizable. In
this regime, conversion of vacuum energy into radiation energy
occurs during the inflationary period and the inflationary regime smoothly
terminates into a radiation dominated regime without an intermediate reheating
period.
The toy model here studied shows that fresh inflation
can be a feasible alternative model to standard inflation\cite{3.}.
The most important characteristic of this model is that shows
a natural transition between the end of inflation and the epoch
when the universe is radiation dominated. 
To conclude, a more exaustive calculation in the inflation/radiation-dominated
interphase should include cold dark matter (CDM)\cite{ul} in this model
of fresh inflation with increasing cosmological parameter.

\noindent
{\bf Fig. 1:} Evolution of the temperature $\theta(t)$ during
inflation. The maximum, with parameters
${\cal M}(0) = 1.5 \times 10^{-9} \  {\rm G}^{-1/2}$, $\lambda = 4\times
10^{-15}$ and $g_{\rm eff}=25$,
occurs at the end of fresh inflation (i.e., around $t_e \simeq 2\times
10^{8} \  {\rm G}^{1/2}$).\\
\vskip 1cm
\noindent
{\bf Fig. 2:} Evolution of density fluctuations
${<\delta\rho_t> \over \rho_t}$ as a function of the number of
$e$-folds, where $N(t_e) \gg 60$. The maximum at the end of inflation
is of the order of $10^{-5}$.\\
\vskip 1cm
\noindent
{\bf Fig. 3:} Temporal evolution of $n$ which shows
that the number of created particles grows with time to
reach the value of $n\simeq 9\times 10^6$ at the end of inflation.\\
\vskip 1cm

\end{document}